\documentclass[11pt]{article}
\usepackage{amssymb,macros2e,graphicx}
\usepackage{times} 
\skip\footins 39pt plus 4pt minus 2pt
\def\footnoterule{\kern-19pt\hrule width.5in\kern18.6pt}%
\newcommand{\dotsb}{\ldots}

\newcommand{\half}{\frac{1}{2}}
\newcommand{\ts}{\hskip0.1ex\raisebox{-1ex}[0ex][0.8ex]{\rule{0.1ex}{2.75ex}\hskip0.2ex}}

\begin{document}
%
\newcommand{\fig}[2]{\includegraphics[width=#1]{./figures/#2}}
\newcommand{\Fig}[1]{\includegraphics[width=\columnwidth]{./figures/#1}}
\newlength{\bilderlength} 
\newcommand{\bilderscale}{0.25}
\newcommand{\storebilderscale}{\bilderscale}
\newcommand{\bilderskip}{\hspace*{0.8ex}}
\newcommand{\textdiagram}[1]{%
\renewcommand{\bilderscale}{0.2}%
\diagram{#1}\renewcommand{\bilderscale}{\storebilderscale}}
\newcommand{\vardiagram}[2]{%
\renewcommand{\bilderscale}{#1}%
\diagram{#2}\renewcommand{\bilderscale}{\storebilderscale}}
\newcommand{\diagram}[1]{%
\settowidth{\bilderlength}{\bilderskip%
\includegraphics[scale=\bilderscale]{./figures/#1}\bilderskip}%
\parbox{\bilderlength}{\bilderskip%
\includegraphics[scale=\bilderscale]{./figures/#1}\bilderskip}}
\newcommand{\Diagram}[1]{%
\settowidth{\bilderlength}{%
\includegraphics[scale=\bilderscale]{./figures/#1}}%
\parbox{\bilderlength}{%
\includegraphics[scale=\bilderscale]{./figures/#1}}}
%

%
\newcommand{\sgn}{{\mathrm{sgn}}}
\newcommand{\rme}{{\mathrm{e}}}
\newcommand{\rmd}{{\mathrm{d}}} 
\newcommand{\nn}{\nonumber}\newcommand {\eq}[1]{(\ref{#1})}
\newcommand {\Eq}[1]{Eq.\hspace{0.55ex}(\ref{#1})}
\newcommand {\Eqs}[1]{Eqs.\hspace{0.55ex}(\ref{#1})}
\newcommand{\E}{\epsilon}
\newcommand{\R}{\mathbb{R}}
\newcommand{\N}{\mathbb{N}}

\def\true{true}
\newsavebox{\bilderbox}
\newlength{\bilderhelp}
\newsavebox{\bilderone}
\newlength{\bilderonelength}
\newsavebox{\bildertwo}
\newlength{\bildertwolength}
%
\newcommand{\bild}[1]{\fboxsep0mm%
\sbox{\bilderbox}{
{\includegraphics[scale=\bilderscale]{#1}}}%
\settowidth{\bilderlength}{\usebox{\bilderbox}}%
\parbox{\bilderlength}{\usebox{\bilderbox}}}
\newcommand{\savebild}[3]{\newsavebox{#2}%
\sbox{#2}{\bild{#3}}\newcommand{#1}{%
\ensuremath{\,\mathchoice{\usebox{#2}}%
{\settowidth{\bilderhelp}{\scalebox{0.7}{\usebox{#2}}}%
\parbox{\bilderhelp}{\scalebox{0.7}{\usebox{#2}}}}%
{\settowidth{\bilderhelp}{\scalebox{0.5}{\usebox{#2}}}%
\parbox{\bilderhelp}{\scalebox{0.5}{\usebox{#2}}}}%
{\settowidth{\bilderhelp}{\scalebox{0.35}{\usebox{#2}}}%
\parbox{\bilderhelp}{\scalebox{0.35}{\usebox{#2}}}}%
\,}}}
\newcommand{\bilderdiagram}[4]{{%
\mathchoice{
\sbox{\bilderone}{\ensuremath{\displaystyle#1}}%
\sbox{\bildertwo}{\ensuremath{\displaystyle#2}}%
\settoheight{\bilderonelength}{\ensuremath{\usebox{\bilderone}}}%
\settoheight{\bildertwolength}{\ensuremath{\usebox{\bildertwo}}}%
\left#3\!\usebox{\bilderone}{\rule{0mm}{\bildertwolength}}\right.%
\hspace*{-0.5ex}\!\left|\!{\rule{0mm}{\bilderonelength}}%
\usebox{\bildertwo}\!\right#4}%
{
\sbox{\bilderone}{\ensuremath{\textstyle#1}}%
\sbox{\bildertwo}{\ensuremath{\textstyle#2}}%
\settoheight{\bilderonelength}{\ensuremath{\usebox{\bilderone}}}%
\settoheight{\bildertwolength}{\ensuremath{\usebox{\bildertwo}}}%
\left#3\!\usebox{\bilderone}{\rule{0mm}{\bildertwolength}}\right.%
\hspace*{-0.35ex}\!\left|\!{\rule{0mm}{\bilderonelength}}%
\usebox{\bildertwo}\!\right#4}%
{
\sbox{\bilderone}{\ensuremath{\scriptstyle#1}}%
\sbox{\bildertwo}{\ensuremath{\scriptstyle#2}}%
\settoheight{\bilderonelength}{\ensuremath{\usebox{\bilderone}}}%
\settoheight{\bildertwolength}{\ensuremath{\usebox{\bildertwo}}}%
\left#3\!\usebox{\bilderone}{\rule{0mm}{\bildertwolength}}\right.%
\hspace*{-0.1ex}\!\left|\!{\rule{0mm}{\bilderonelength}}%
\usebox{\bildertwo}\!\right#4}%
{
\sbox{\bilderone}{\ensuremath{\scriptscriptstyle#1}}%
\sbox{\bildertwo}{\ensuremath{\scriptscriptstyle#2}}%
\settoheight{\bilderonelength}{\ensuremath{\usebox{\bilderone}}}%
\settoheight{\bildertwolength}{\ensuremath{\usebox{\bildertwo}}}%
\left#3\!\usebox{\bilderone}{\rule{0mm}{\bildertwolength}}\right.%
\hspace*{-0.1ex}\!\left|\!{\rule{0mm}{\bilderonelength}}%
\usebox{\bildertwo}\!\right#4}%
}}
\newcommand{\MOPE}[2]{\bilderdiagram{#1}{#2}{(}{)}}
\newcommand{\DIAG}[2]{\bilderdiagram{#1}{#2}{<}{>}}
\newcommand{\DIAGindhelp}[4]{%
\sbox{\bilderbox}{\ensuremath{#4\bilderdiagram{#1}{#2}{<}{>}}}%
\settowidth{\bilderlength}{\rotatebox{90}{\ensuremath{\usebox{\bilderbox}}}}%
\ensuremath{\usebox{\bilderbox}_{\hspace*{-0.162\bilderlength}#3}}}
\newcommand{\DIAGind}[3]{%
\mathchoice{\DIAGindhelp{#1}{#2}{#3}{\displaystyle}}%
{\DIAGindhelp{#1}{#2}{#3}{\textstyle}}%
{\DIAGindhelp{#1}{#2}{#3}{\scriptstyle}}%
{\DIAGindhelp{#1}{#2}{#3}{\scriptscriptstyle}}}
\newcommand{\reducedbildheightrule}[2]{{%
\mathchoice{\settowidth{\bilderlength}{\rotatebox{90}{\ensuremath{\displaystyle#1}}}%
\parbox{0mm}{\rule{0mm}{#2\bilderlength}}}%
{\settowidth{\bilderlength}{\rotatebox{90}{\ensuremath{\textstyle#1}}}%
\parbox{0mm}{\rule{0mm}{#2\bilderlength}}}%
{\settowidth{\bilderlength}{\rotatebox{90}{\ensuremath{\scriptstyle#1}}}%
\parbox{0mm}{\rule{0mm}{#2\bilderlength}}}%
{\settowidth{\bilderlength}{\rotatebox{90}{\ensuremath{\scriptscriptstyle#1}}}%
\parbox{0mm}{\rule{0mm}{#2\bilderlength}}}}}
\newcommand{\bildheightrule}[1]{\reducedbildheightrule{#1}{1}}
\newcommand{\reducedbild}[2]{%
{\settowidth{\bilderhelp}{#1}%
\setlength{\bilderhelp}{#2\bilderhelp}%
\parbox{\bilderhelp}{\scalebox{#2}{#1}}}}
%
%

\centerline{\sffamily\bfseries\large  The Functional Renormalization
Group Treatment of Disordered Systems,}\smallskip

\centerline{\sffamily\bfseries\large
a Review} \smallskip

\bigskip  
\centerline{\sffamily\bfseries\normalsize Kay J\"org Wiese}
\bigskip 
\centerline{KITP, Kohn Hall, University of
California at 
Santa Barbara, Santa Barbara, CA 93106, USA}
\medskip 

\medskip 
\centerline{\small Nov.~23, 2002}

\noindent \rule{\textwidth}{0.3mm} \smallskip \leftline{\bfseries
Abstract} We review current progress in the functional renormalization
group treatment of disordered systems. After an elementary
introduction into the phenomenology, we show why in the context of
disordered systems a functional renormalization group treatment is
necessary, contrary to pure systems, where renormalization of a single
coupling constant is sufficient. This leads to a disorder
distribution, which after a finite renomalization becomes
non-analytic, thus overcoming the predictions of the seemingly exact
dimensional reduction.  We discuss, how a renormalizable field theory
can be constructed, even beyond 1-loop order. We then discuss an
elastic manifold imbedded in $N$ dimensions, and give the exact
solution for $N\to\infty$. This is compared to predictions of the
Gaussian replica variational ansatz, using replica symmetry
breaking. We finally discuss depinning, both isotropic and
anisotropic, and the scaling function for the width distribution of an
interface.

\noindent\rule{\textwidth}{0.3mm}

{\small\tableofcontents}
\newpage


\section{Introduction}\label{intro}

Statistical mechanics is by now a rather mature branch of physics.
For pure systems like a ferromagnet, it allows to calculate so precise
details as the behavior of the specific heat on approaching the
Curie-point. We know that it diverges as a function of the distance in
temperature to the Curie-temperature, we know that this divergence has
the form of a power-law, we can calculate the exponent, and we can do
this with at least 3 digits of accuracy. Best of all, these findings
are in excellent agreement with the most precise experiments. This is
a true success story of statistical mechanics.  On the other hand, in
nature no system is really pure, i.e.\ without at least some disorder
(``dirt'').  As experiments (and theory) seem to suggest, a little bit
of disorder does not change the behavior much. Otherwise experiments
on the specific heat of Helium would not so extraordinarily well
confirm theoretical predictions. But what happens for strong disorder?
By this I mean that disorder completely dominates over entropy. Then
already the question: ``What is the ground-state?'' is no longer
simple. This goes hand in hand with the appearance of so-called
metastable states. States, which in energy are very close to the
ground-state, but which in configuration-space may be far apart. Any
relaxational dynamics will take an enormous time to find the correct
ground-state, and may fail altogether, as can be seen in
computer-simulations as well as in experiments. This means that our
way of thinking, taught in the treatment of pure systems, has to be
adapted to account for disorder. We will see that in contrast to pure
systems, whose universal large-scale properties can be described by
very few parameters, disordered systems demand the knowledge of the
whole disorder-distribution function (in contrast to its first few
moments). We show how universality nevertheless emerges.

Experimental realizations of strongly disordered systems are glasses,
or more specifically spin-glasses, vortex-glasses, electron-glasses
and structural glasses (not treated here).  Furthermore random-field
magnets, and last not least elastic systems in disorder.

What is our current understanding of disordered systems? It is here
that the success story of statistical mechanics, with which I started,
comes to an end: Despite 30 years of research, we do not know much:
There are a few exact solutions, there are phenomenological methods
(like the droplet-model), and there is the mean-field approximation,
involving a method called replica-symmetry breaking (RSB). This method
is correct for infinitely connected systems, e.g.\ the SK-model
(Sherrington Kirkpatrick model), or for systems with infinitely many
components.  However it is unclear, to which extend it applies to real
physical systems, in which each degree of freedom is only coupled to a
finite number of other degrees of freedom.

Another interesting system are elastic manifolds in a random media,
which has the advantage of being approachable by other (analytic)
methods, while still retaining all the rich physics of strongly
disordered systems.  Here, I review recent advances obtained in
collaboration with Pierre Le Doussal 
\cite{ChauveLeDoussalWiese2000a,LeDoussalWiese2001,LeDoussalWieseChauve2002,LeDoussalWieseChauvePREPa,LeDoussalWiese2002a,LeDoussalWiesePREPb,LeDoussalWiesePREPc,LeDoussalWiesePREPd,LeDoussalWiesePREPf,LeDoussalWiesePREPg,LeDoussalWiese2003a,RossoKrauthLeDoussalVannimenusWiese2003}.
This review is an extended version of \cite{Wiese2002}.

\section{Physical realizations, model and observables}\label{model}
\begin{figure}[t]
\centerline{\fig{0.25\textwidth}{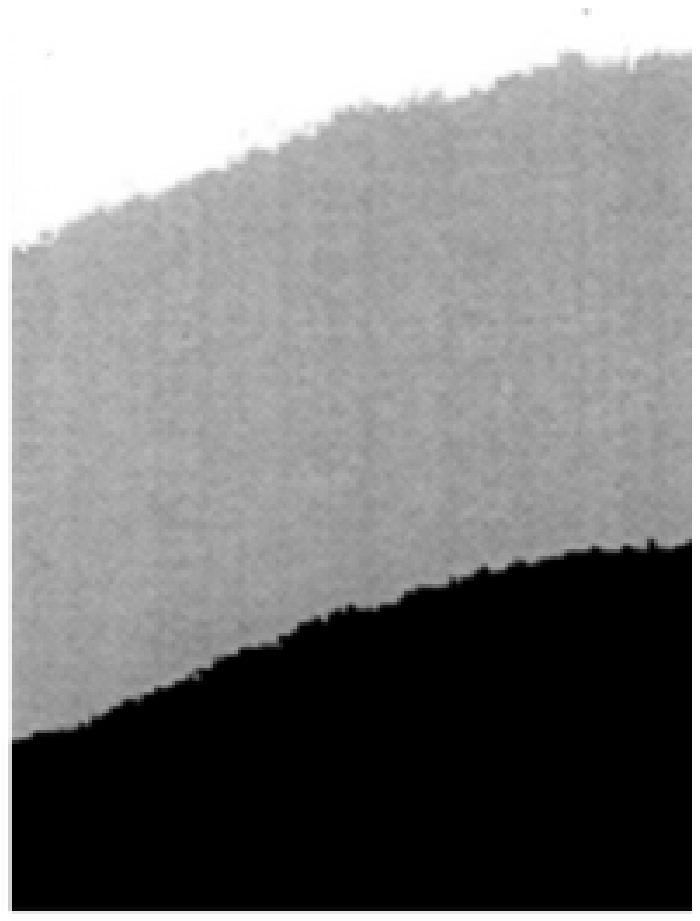}~~~\fig{0.7\textwidth}{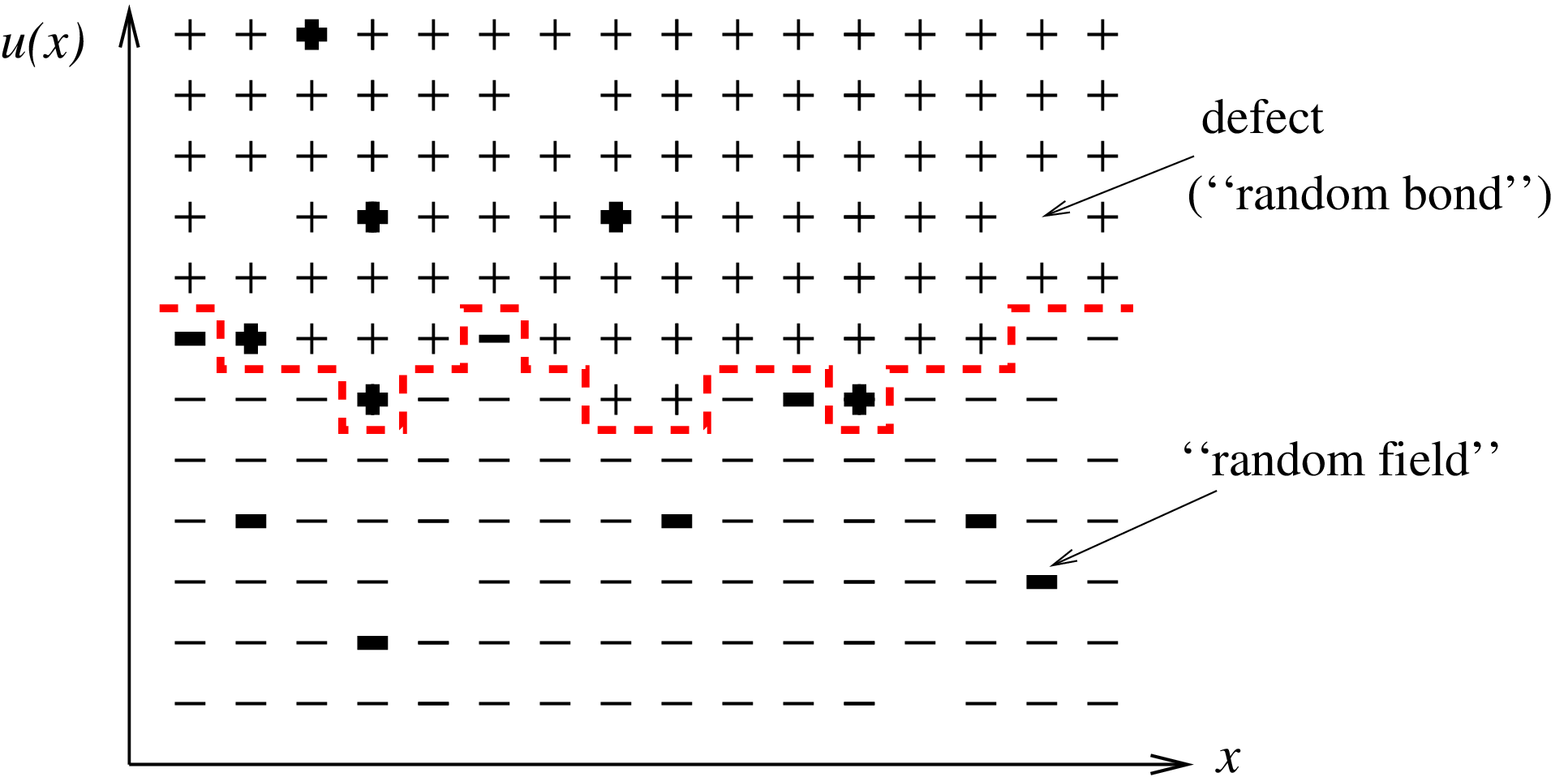}}
\caption{An Ising magnet at low temperatures forms a domain wall
described by a function $u (x)$ (right). An experiment on a thin
Cobalt film (left)
\protect\cite{LemerleFerreChappertMathetGiamarchiLeDoussal1998}; with
kind permission of the authors.}
\label{exp:Magnet}
\end{figure}
\begin{figure}[b]
\centerline{\parbox{0.5\textwidth}{\fig{0.5\textwidth}{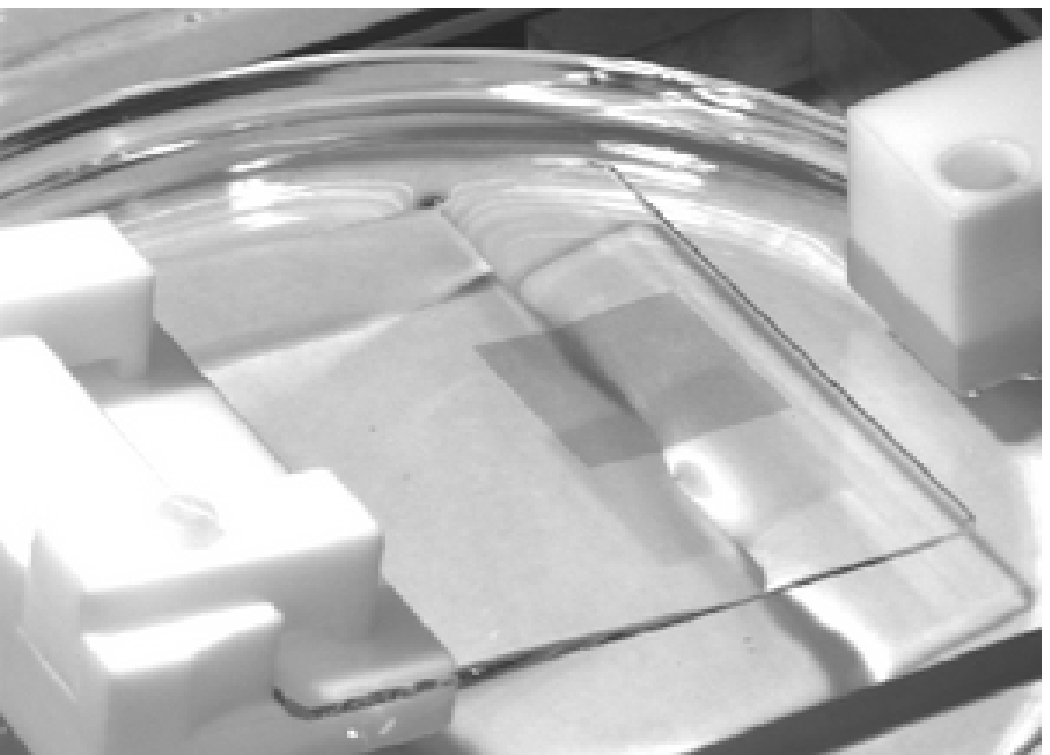}}
\parbox{0.445\textwidth}{\begin{minipage}{0.445\textwidth}
\Fig{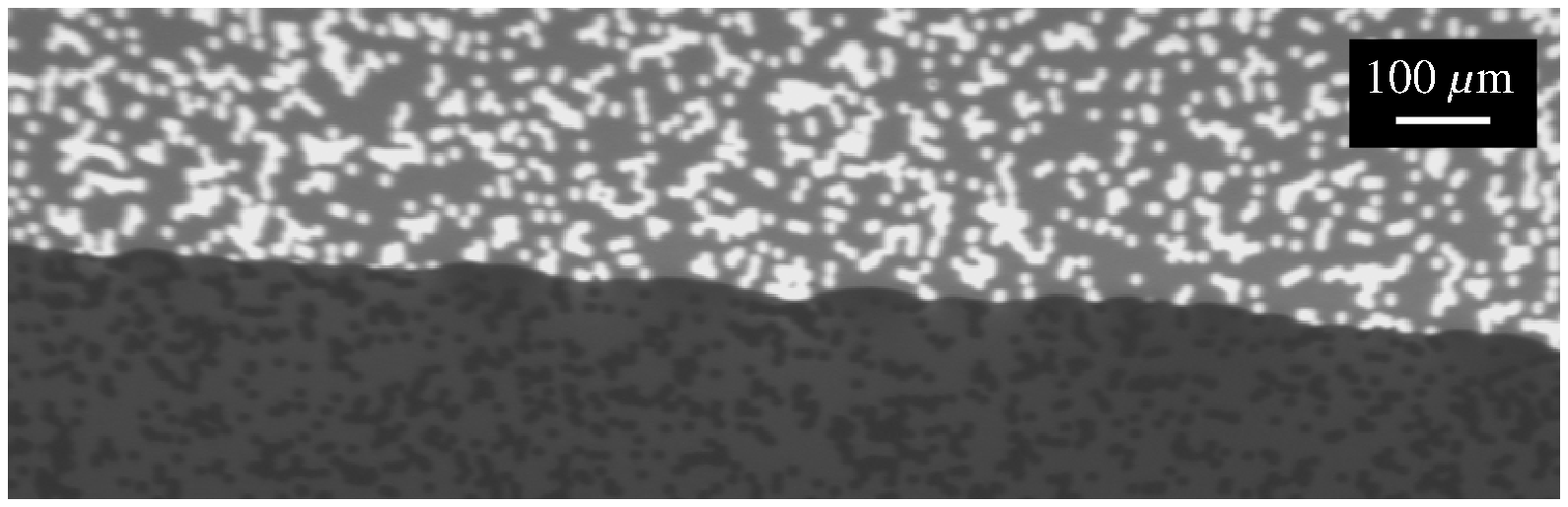}\\
\Fig{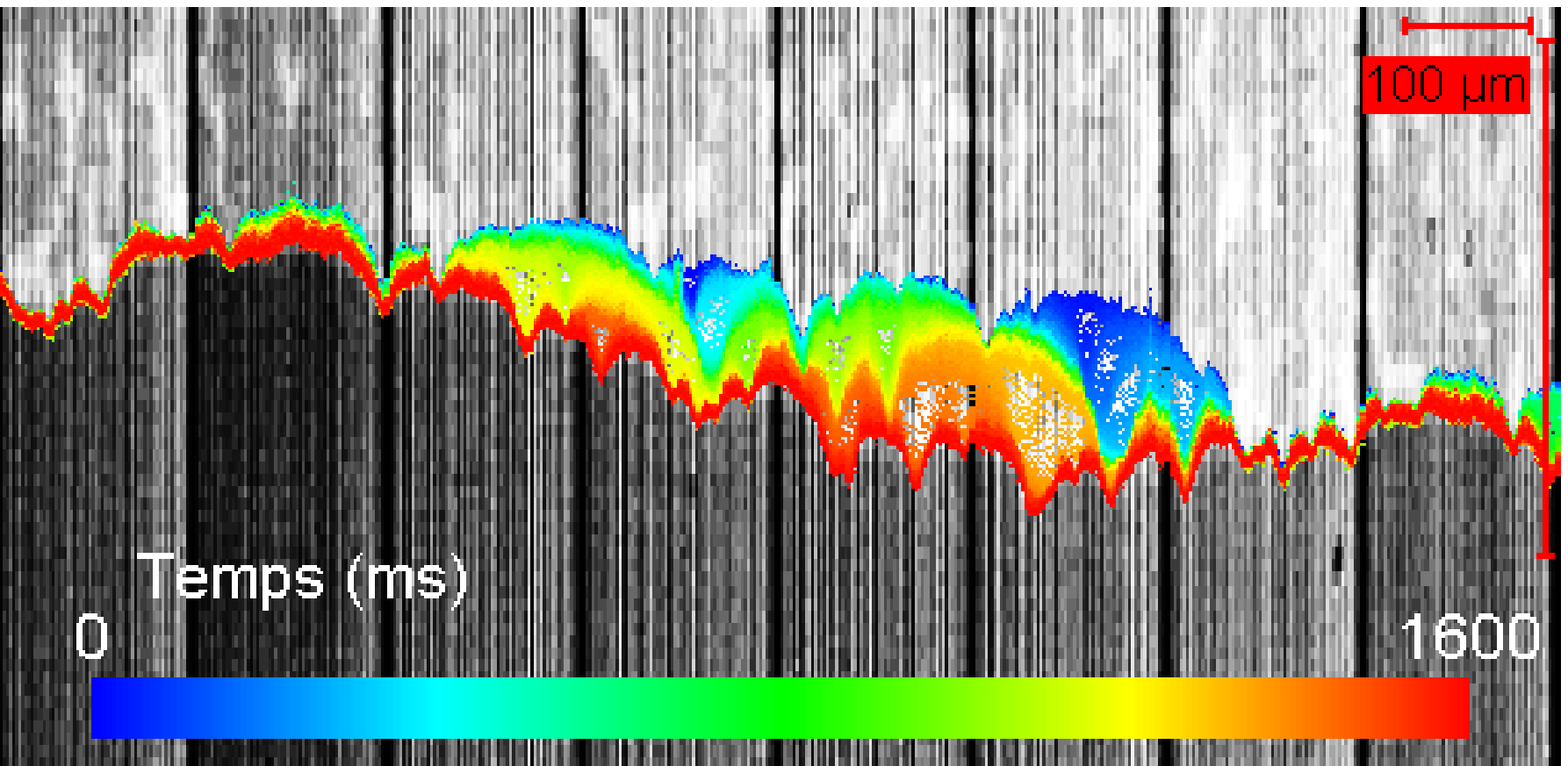}
\end{minipage}}}
\caption{A contact line for the wetting of a disordered substrate by
Glycerine \protect\cite{MoulinetGuthmannRolley2002}. Experimental setup
(left). The disorder consists of randomly deposited islands of
Chromium, appearing as bright spots (top right). Temporal evolution of
the retreating contact-line (bottom right). Note the different scales
parallel and perpendicular to the contact-line. Pictures courtesy of
S.~Moulinet, with kind permission.}  \label{exp:contact-line}
\end{figure}
\begin{figure}[t]\label{f:vortex-lattic}
\centerline{\parbox{0.47\textwidth}{\fig{0.47\textwidth}{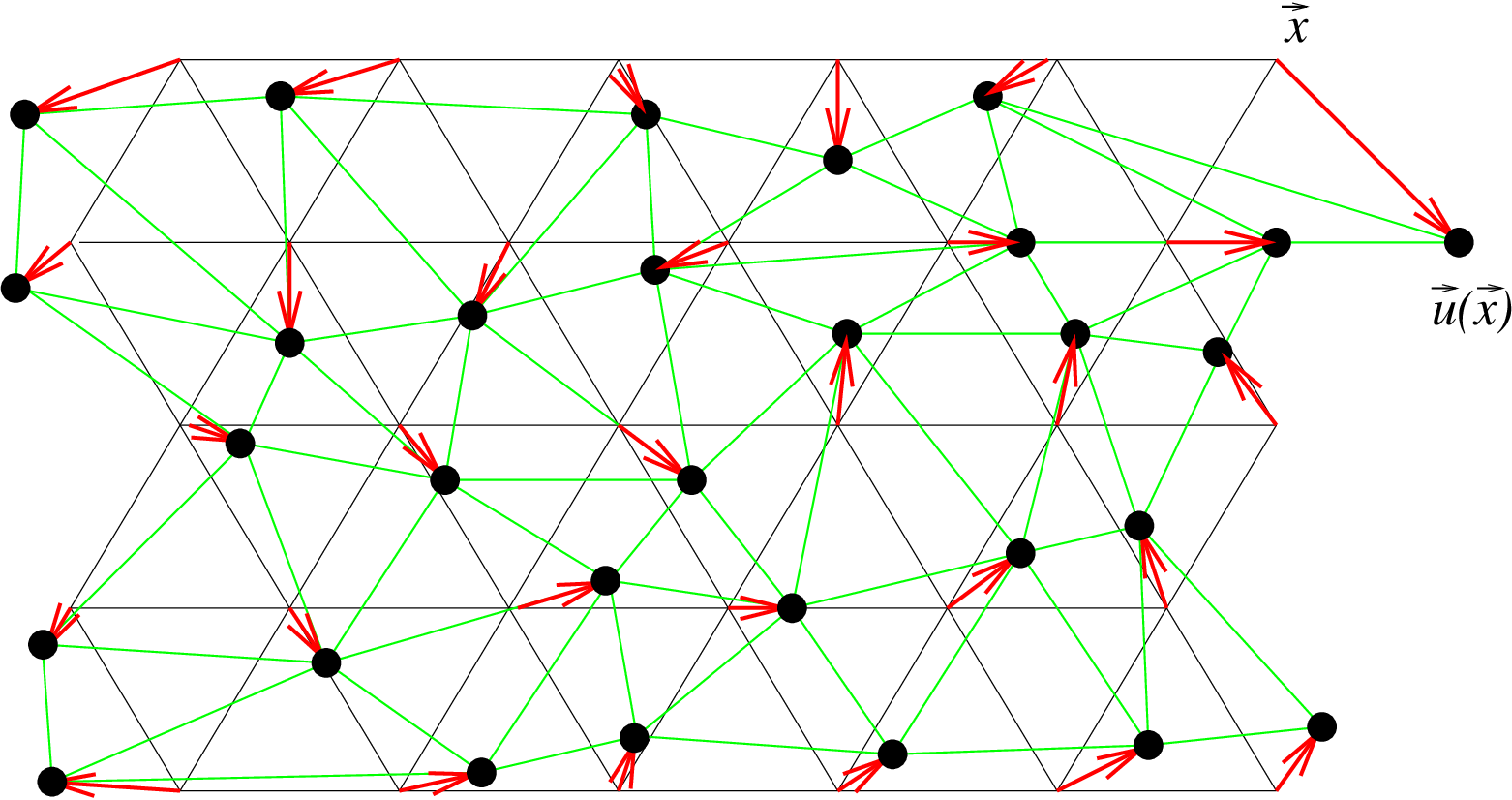}}}\smallskip
\caption{Cartoon of an  elastic lattice (e.g.\ vortex lattice)
deformed by disorder. This is described by a vector $\vec u
(x)$.} 
\end{figure}

Before developing the theory to treat elastic systems in a disordered
environment, let us give some physical realizations. The simplest one
is an Ising magnet. Imposing boundary conditions with all spins up at
the upper and all spins down at the lower boundary (see figure 1), at
low temperatures, a domain wall separates a region with spin up from a
region with spin down. In a pure system at temperature $T=0$, this
domain wall is completely flat.  Disorder can deform the domain wall,
making it eventually rough again. Two types of disorder are common:
random bond (which on a course-grained level also represents missing
spins) and random field (coupling of the spins to an external random
magnetic field). Figure 1 shows, how the domain wall is described by a
displacement field $u (x)$.  Another example is the contact line of
water (or liquid Helium), wetting a rough substrate, see figure
\ref{exp:contact-line}. (The elasticity is long range). A realization
with a 2-parameter displacement field $\vec{u} (\vec x) $ is the
deformation of a vortex lattice: the position of each vortex is
deformed from $\vec x$ to $\vec x+ \vec u (\vec x)$.  A 3-dimensional
example are charge density waves.

All these models have in common, that they are described 
by a displacement field 
\begin{equation}\label{u}
x\in \R^d \ \longrightarrow\  \vec u (x) \in \R^N
\ .
\end{equation}
For simplicity, we set $N=1$ in the following.  After some initial
coarse-graining, the energy ${\cal H}={\cal H}_{\mathrm{el}}+{\cal
H}_{\mathrm{DO}}$ consists out of two parts: the elastic energy
\begin{equation}
{\cal H}_{\mathrm{el}}[u] = \int \rmd ^d x \, \half \left( \nabla u
(x)\right)^2 
\end{equation}
and the disorder
\begin{equation}
{\cal H}_{\mathrm{DO}}[u] = \int \rmd ^{d} x \, V (x,u (x))\ .
\end{equation}
In order to proceed, we need to specify the  correlations of
disorder. Suppose that fluctuations $u$ in
the transversal direction scale  as
\begin{equation}\label{roughness}
\overline{\left(u (x)-u (y) \right)^{2}}  \sim  |x-y|^{2\zeta }
\end{equation}
with a roughness-exponent $\zeta <1$. Starting from a disorder
correlator 
\begin{equation}
\overline{V (u,x)V (u',x')} = f (x-x') R (u-u')
\end{equation}
and performing one step in the RG-procedure, one has to rescale more
in the $x$-direction than in the $u$-direction. This will eventually
reduce $f (x-x')$ to a $\delta $-distribution, whereas the structure
of $R (u-u')$ remains visible.  We therefore choose as our
starting-model
\begin{equation}\label{DOcorrelR}
\overline{V (u,x)V (u',x')} := \delta ^{d } (x-x') R (u-u')
\ .
\end{equation}
There are a couple of useful observables. We already mentioned the
roughness-exponent $\zeta $. The second is the renormalized
(effective) disorder. It will turn out that we actually have to keep
the whole disorder distribution function $R (u)$, in contrast to
keeping a few moments.  Other observables are higher correlation
functions or the free energy.

\section{Treatment of disorder}\label{treat disorder} Having defined
our model, we can now turn to the treatment of disorder. The problem
is to average not the partition-function, but the free energy over
disorder: $\overline{{\cal F}}=\overline{\ln Z} $. This can be
achieved by the beautiful {\em replica-trick}. The idea is to write
\begin{equation}
\ln {\cal Z} = \lim_{n\to 0} \frac{1}{n}\left( \rme^{n \ln {\cal Z}}-1
\right) = \lim_{n\to 0} \frac{1}{n}\left({\cal Z}^{n}-1 \right)
\end{equation}
and to interpret ${\cal Z}^{n}$ as the partition-function of an $n$
times replicated system. Averaging $\rme ^{-\sum _{a=1}^{n}{\cal
H}_{a}}$ over disorder then leads to the {\em replica-Hamiltonian}
\begin{equation}\label{H}
{\cal H}[u] = \frac{1}{T} \sum _{a=1}^{n}\int \rmd ^{d }x\, \half
\left(\nabla u_{a} (x) \right)^{2} -\frac{1}{2 T^{2}}  \sum
_{a,b=1}^{n} \int \rmd ^{d }x\, R (u_{a} (x)-u_{b} (x))\ .
\end{equation}
Let us stress that one could equivalently pursue a dynamic or a
supersymmetric formulation. We therefore should not, and in fact do
not encounter, problems associated with the use of the replica-trick.

\section{Dimensional reduction} There is a beautiful and rather
mind-boggling theorem relating disordered systems to pure systems
(i.e.\ without disorder), which applies to a large class of systems,
e.g.\ random field systems and elastic manifolds in disorder. It is
called dimensional reduction and reads as
follows\cite{EfetovLarkin1977}:

\noindent {\underline{Theorem:}} {\em A $d$-dimensional disordered
system at zero temperature is equivalent to all orders in perturbation
theory to a pure system in $d-2$ dimensions at finite temperature. }
Moreover the temperature is (up to a constant) nothing but the width
of the disorder distribution. A simple example is the 3-dimensional
random-field Ising model at zero temperature; according to the theorem
it should be equivalent to the pure 1-dimensional Ising-model at
finite temperature. But it has been shown rigorously, that the former
has an ordered phase, whereas we have all solved the latter and we
know that there is no such phase at finite temperature. So what went
wrong? Let me stress that there are no missing diagrams or any such
thing, but that the problem is more fundamental: As we will see later,
the proof makes assumptions, which are not satisfied.  Nevertheless,
the above theorem remains important since it has a devastating
consequence for all perturbative calculations in the disorder: However
clever a procedure we invent, as long as we do a perturbative
expansion, expanding the disorder in its moments, all our efforts are
futile: dimensional reduction tells us that we get a trivial and
unphysical result. Before we try to understand why this is so and how
to overcome it, let me give one more example. Dimensional reduction
allows to calculate the roughness-exponent $\zeta $ defined in
equation (\ref{roughness}).  We know (this can be inferred from
power-counting) that the width $u$ of a $d$-dimensional manifold at
finite temperature in the absence of disorder scales as $u\sim
x^{(2-d)/2}$. Making the dimensional shift implied by dimensional
reduction leads to
\begin{equation}\label{zetaDR}
\overline{\left( u (x)-u (0) \right)^{2}} \sim x^{4-d} \equiv x^{2\zeta }
\quad \mbox{i.e.}\quad \zeta =\frac{4-d}{2}\ .
\end{equation}

\section{The Larkin-length}\label{Larkin} To understand the failure of
dimensional reduction, let us turn to an interesting argument given by
Larkin \cite{Larkin1970}. He considers a piece of an elastic manifold
of size $L$. If the disorder has correlation length $r$, and
characteristic potential energy $\bar f$, this piece will typically
see a potential energy of strength
\begin{equation}
E_{\mathrm{DO}} = \bar f \left(\frac{L}{r} \right)^{\!\frac{d}{2}}\ . 
\end{equation}
On the other hand, there is an elastic energy, which scales like 
\begin{equation}
E_{\mathrm{el}} = c\, L^{d-2}\ . 
\end{equation}
These energies are balanced at the  {\em Larkin-length} $L=L_{c}$
with 
\begin{equation}
L_{c} = \left(\frac{c^{2}}{\bar f^{2}}r^{d} \right)^{\frac{1}{4-d}}
\ .
\end{equation}
More important than this value is the observation that in all
physically interesting dimensions $d<4$, and at scales $L>L_{c}$, the
membrane is pinned by disorder; whereas on small scales elastic energy
dominates. Since the disorder has a lot of minima which are far apart
in configurational space but close in energy (metastability), the
manifold can be in either of these minimas, and the ground-state is no
longer unique. However exactly this is assumed in e.g.\ the proof of
dimensional reduction; as is most easily seen in its supersymmetric
formulation \cite{ParisiSourlas1979}.

\section{The functional renormalization group (FRG)}\label{FRG}
\begin{figure}[t]\begin{center}
{\unitlength1.0mm\fboxsep0mm 
\mbox{\begin{picture} (134,30)
\put(0,25){$R (u_{a} (x)-u_{b} (x))$}
\put(5,20){$ = \diagram{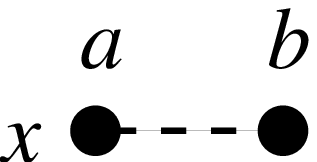}$}
\put(0,10){$C (x-y)=$}
\put(0,5) {$\ \diagram{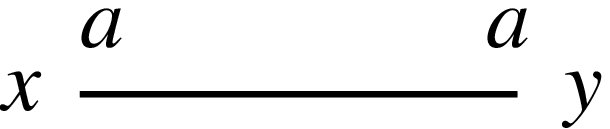}$}
\put(40,19){$\delta R (u_{a}-u_{b}) =
\diagram{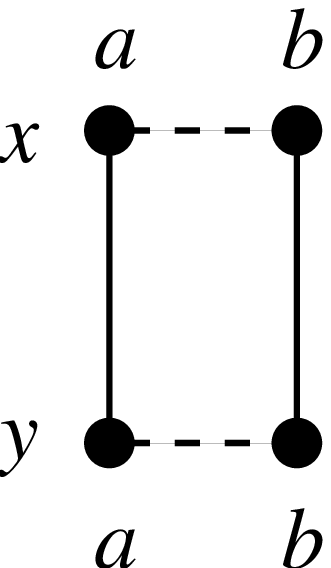}-2\diagram{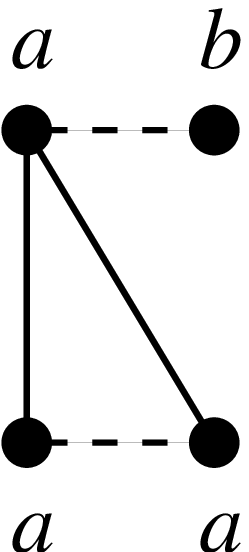} $}
\put(40,4){$=\displaystyle \int_{x-y}C (x-y)^{2} \left[  R'' (u_{a}
-u_{b})^{2} - 2 R'' (u_{a}-u_{b})R'' (u_{a}-u_{a})\right]$} 
\end{picture}} }
\end{center}
\caption{The disorder vertex $R (u_{a} (x)-u_{b} (x))$ and the correlation-%
function $C (x-y)$, with Fourier-transform $\tilde C
(k)=\frac{1}{k^{2}}$, which is diagonal in replica-space
(left). Contracting two disorder-vertices with two 
correlation-functions leads to the two 1-loop contributions $\delta R$
to the disorder-correlator $R$ (right). The integral $\int_{x-y}C
(x-y)^{2}=\frac{L^{\epsilon }}{\epsilon }$, where $L$ is some
IR-cutoff.}\label{fig:RG}
\end{figure}
Let us now discuss a way out of the dilemma: Larkin's argument
suggests that $4$ is the upper critical dimension. So we would like to
make an $\epsilon =4-d$ expansion. On the other hand, dimensional
reduction tells us that the roughness is $\zeta =\frac{4-d}{2}$ (see
(\ref{zetaDR})). Even though this is systematically wrong below four
dimensions, it tells us correctly that at the critical dimension
$d=4$, where disorder is marginally relevant, the field $u$ is
dimensionless. This means that having identified any relevant or
marginal perturbation (as the disorder), we find immediately another
such perturbation by adding more powers of the field. We can thus not
restrict ourselves to keeping solely the first moments of the
disorder, but have to keep the whole disorder-distribution function $R
(u)$. Thus we need a {\em functional renormalization group} treatment
(FRG). Functional renormalization is an old idea going back to the
seventies, and can e.g.\ be found in \cite{WegnerHoughton1973}.  For
disordered systems, it was first proposed in 1986 by D.\ Fisher
\cite{DSFisher1986}.  Performing an infinitesimal renormalization,
i.e.\ integrating over a momentum shell \`a la Wilson, leads to the
flow $\partial _{\ell} R (u)$, with ($\epsilon =4-d$)
\begin{equation}\label{1loopRG}
\partial _{\ell} R (u) = \left(\epsilon -4 \zeta  \right) R (u) +
\zeta u R' (u) + \frac{1}{2} R'' (u)^{2}-R'' (u)R'' (0)\ . 
\end{equation}
The first two terms come from the rescaling of $R$ and $u$
respectively. The last two terms are the result of the 1-loop
calculations, which are sketched in figure \ref{fig:RG}.

\begin{figure}[t]
\centerline{\fig{13.4cm}{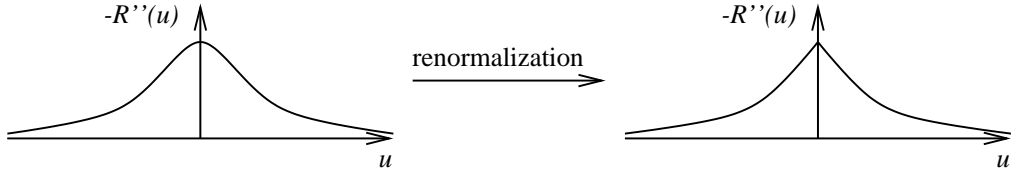}}
\caption{Change of $-R'' (u)$ under renormalization and formation of
the cusp.} \label{fig:cusp}
\end{figure}

More important than the form of this equation is it actual solution,
sketched in figure \ref{fig:cusp}.
After some finite renormalization, the second derivative of the
disorder $R'' (u)$ acquires a cusp at $u=0$; the length at which this
happens is the Larkin-length. How does this overcome dimensional
reduction?  To understand this, it is interesting to study the flow of
the second and forth moment. Taking derivatives of (\ref{1loopRG})
w.r.t.\ $u$ and setting $u$ to 0, we obtain
\begin{eqnarray}
\partial_{\ell} R'' (0) &=& \left(\epsilon -2 \zeta  \right) R'' (0) +
R''' (0)^{2} \ \longrightarrow \ \left(\epsilon -2 \zeta  \right) R''
(0)\label{R2of0}\\ 
\partial_{\ell} R'''' (0) &=& \epsilon  R'''' (0) + 3 R'''' (0)^{2} +4 R'''
(0)R''''' (0)  \ \longrightarrow\ \epsilon  R'''' (0) + 3 R''''
(0)^{2}\label{R4of0} 
\ .
\end{eqnarray}
Since $R (u)$ is an even function, $R''' (0)$ and $R''''' (0)$ are 0,
which we have already indicated in Eqs.\ (\ref{R2of0}) and
(\ref{R4of0}) . The above equations for $R'' (0)$ and $R'''' (0)$ are
in fact closed.  Equation (\ref{R2of0}) tells us that the flow of $R''
(0)$ is trivial and that $\zeta =\epsilon /2\equiv
\frac{4-d}{2}$. This is exactly the result predicted by dimensional
reduction. The appearance of the cusp can be inferred from equation
(\ref{R4of0}). Its solution is
\begin{equation}
R'''' (0)\ts _{\ell}= \frac{c\,\rme^ {\epsilon \ell }}{1-3\, c \left(\rme^
{\epsilon \ell} -1 \right)/ \epsilon }\ , \qquad c:= R'''' (0)\ts _{\ell=0}
\end{equation}
Thus after a finite renormalization $R'''' (0)$ becomes infinite: The
cusp appears. By analyzing the solution of the flow-equation
(\ref{1loopRG}), one also finds that beyond the Larkin-length $R''
(0)$ is no longer given by (\ref{R2of0}) with $R''' (0)^{2}=0$.  The
correct interpretation of (\ref{R2of0}), which remains valid after the
cusp-formation, is (for details see below)
\begin{equation}
\partial_{\ell} R'' (0) = \left(\epsilon -2 \zeta  \right) R'' (0)  +R''' (0^{+})^{2} \label{R2of0after}\ .
\end{equation} 
Renormalization of the whole function thus overcomes dimensional
reduction.  The appearance of the cusp also explains why dimensional
reduction breaks down. The simplest way to see this is by redoing the
proof for elastic manifolds in disorder, which in the absence of
disorder is a simple Gaussian theory. Terms contributing to the
2-point function involve $R'' (0)$, $TR'''' (0)$ and higher
derivatives of $R (u)$ at $u=0$, which all come with higher powers of
$T$. To obtain the limit of $T\to 0$, one sets $T=0$, and only $R''
(0)$ remains. This is the dimensional reduction result. However we
just saw that $R'''' (0)$ becomes infinite. Thus $R'''' (0) T$ may
also contribute, and the proof fails.

\section{Why is a cusp necessary?}  
\begin{figure}[t] \begin{center}{\unitlength1mm\fboxsep0mm
\mbox{\begin{picture} (134,20)
\put(17,16.5){${\cal H}[u]$}
\put(52,1){$u$}
\put(98,16.5){${\cal H}[u]$}
\put(60,13.7){$L<L_{c}$} 
\put(61,5.5){$L>L_{c}$} 
\put(132.5,1){$u$}
\put(0,1){\fig{132mm}{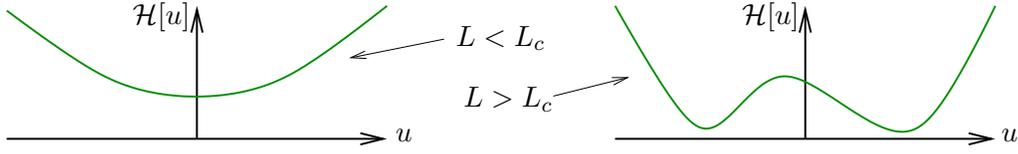} } 
\end{picture} }}
\end{center}
\caption{The toy model (\ref{toy}) before (left) and after (right) the
Larkin-scale.}\label{fig:toy}
\end{figure}

The appearance of a cusp might
suggest that this approach is fatally ill. Let me present a simple
argument due to Leon Balents \cite{BalentsPrivate}, why a cusp {\em is
a physical necessity and not an artifact.} To this aim, consider a toy
model with only one Fourier-mode $u=u_{q}$
\begin{equation}\label{toy}
{\cal H}[u] = \half q^{2 } u^{2} + \sqrt{\epsilon }\, \tilde V (u)
\ .
\end{equation}
Since equation (\ref{1loopRG}) has a fixed point of order $R (u)\sim
\epsilon $ for all $\epsilon >0$, $V (u)$ scales like $\sqrt{\epsilon
}$ for $\epsilon $ small and we have made this dependence explicit in
(\ref{toy}) by using $V (u)= \sqrt{\epsilon }\tilde V (u)$. The only
further input comes from the physics: For $L<L_{c}$, i.e.\ before we
reach the Larkin length, there is only one minimum, as depicted in
figure \ref{fig:toy}. On the other hand, for $L>L_{c}$, there are
several minima. Thus there is at least one point for which
\begin{equation}
\frac{\rmd ^2}{\rmd u^2}\, {\cal H}[u] = q^{2 } + \sqrt{\epsilon }\, \tilde V''
(u) < 0
\ .
\end{equation}
In the limit of $\epsilon \to 0$, this is possible if and only if
$\frac{1}{\epsilon }R'''' (0)$, which a priori should be finite for
$\epsilon \to 0$, becomes infinite:
\begin{equation}
\frac{1}{\epsilon }R'''' (0) = \overline{V'' (u)V'' (u')}\ts _{u=u'} = \infty 
\ .
\end{equation}
This argument shows that a cusp is indeed a physical necessity.

\section{Beyond 1 loop?}\label{beyond1loop} Functional renormalization
has successfully been applied to a bunch of problems at 1-loop
order. From a field theory, we however demand more. Namely that
it\medskip

$\bullet$ allows for systematic corrections beyond 1-loop order\smallskip

$\bullet$ be renormalizable\smallskip

$\bullet$ and thus allows to make universal predictions.\medskip

\noindent However, this has been a puzzle since 1986, and it has even
been suggested that the theory is not renormalizable due to the
appearance of terms of order $\epsilon ^{\frac{3}{2}}$
\cite{BalentsDSFisher1993}. Why is the next order so complicated? The
reason is that it involves terms proportional to $R''' (0)$. A look at
figure 3 explains the puzzle. Shall we use the symmetry of $R (u)$ to
conclude that $R''' (0)$ is 0? Or shall we take the left-hand or
right-hand derivatives, related by
\begin{equation}
R''' (0^{+}) := \lim_{{u>0}\atop {u\to 0}} R ''' (u) = -
\lim_{{u<0}\atop {u\to 0}} R ''' (u) =:- R''' (0^{-}) .
\end{equation}
In the following, I will present our solution of this puzzle, obtained
at 2-loop order in \cite{ChauveLeDoussalWiese2000a} and then at large
$N$ in \cite{LeDoussalWiese2001}. The latter approach allows for
another independent control-parameter, and sheds further light on the
cusp-formation. We refer the reader to these works for further
references.

\section{Results at 2-loop order}\label{2loop}
For the flow-equation at 2-loop order, we find
\cite{ChauveLeDoussalWiese2000a}  
\begin{eqnarray}\label{2loopRG}
\partial _{\ell} R (u) &=& \left(\epsilon -4 \zeta  \right) R (u) +
\zeta u R' (u) + \frac{1}{2} R'' (u)^{2}-R'' (u)R'' (0) \nn \\
&& + \frac{1}{2}\left(R'' (u)-R'' (0) \right)R'''
(u)^{2}-\frac{1}{2}R''' (0^{+})^{2 } R'' (u) \ .
\end{eqnarray}
The first line is the result at 1-loop order, already given in
(\ref{1loopRG}). The second line is new. The most interesting term is
the last one, which involves $R''' (0^{+})^{2}$ and which we therefore
call {\em anomalous}.  The hard task is to fix the prefactor
$(-\frac{1}{2})$.  We have up to now invented six algorithms to do so;
one leads to inconsistencies and shall not be reported here. The other
five algorithms are consistent with each other: The sloop-algorithm,
recursive construction, reparametrization invariance,
renormalizability, and potentiality
\cite{ChauveLeDoussalWiese2000a,LeDoussalWieseChauvePREPa}. For lack
of space, we restrain our discussion to the last two ones. At 2-loop
order the following diagram appears
\begin{equation}\label{rebi}
\diagram{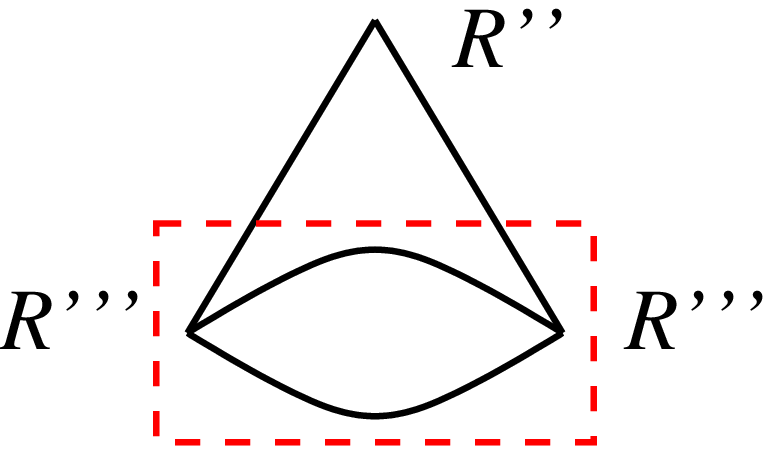}\  \longrightarrow\ \frac{1}{2}\left(R'' (u)-R'' (0)
\right)R''' 
(u)^{2} -\half R'' (u)R''' (0^{+})^{2}
\end{equation}
leading to the anomalous term. The integral (not written here)
contains a subdivergence, which is indicated by the
box. Renormalizability demands that its leading divergence (which is
of order $1/\epsilon ^{2}$) be canceled by a 1-loop counter-term. The
latter is unique thus fixing the prefactor of the anomalous term. (The
idea is to take the 1-loop correction $\delta R$ in figure 2 and
replace one of the $R''$ in it by $\delta R''$ itself, which the
reader can check to leading to the terms given in (\ref{rebi}) plus
terms which only involve even derivatives.)

Another very physical demand is that the problem remain potential,
i.e.\ that forces still derive from a potential. The force-force
correlation function being $-R'' (u)$, this means that the flow of
$R' (0)$ has to be strictly 0. (The simplest way to see this is to
study a periodic potential.) From (\ref{2loop}) one can check that
this does not remain true if one changes the prefactor of the last
term in (\ref{2loop}); thus fixing it.

Let us give some results for cases of physical interest. First of all,
in the case of a periodic potential, which is relevant for
charge-density waves, the fixed-point function can be calculated
analytically as (we choose period 1, the following is for $u\in
\left[0,1 \right]$)
\begin{equation}
R^{*} (u) = - \left(\frac{\epsilon }{72}+\frac{\epsilon ^{2}}{108}+O
(\epsilon ^{3}) \right) u^{2} (1-u)^{2} +\mbox{const.}
\end{equation}
This leads to a universal amplitude.  In the case of random field
disorder (short-ranged force-force correlation function) $\zeta
=\frac{\epsilon }{3}$. For random-bond disorder (short-ranged
potential-potential correlation function) we have to solve
(\ref{2loopRG}) numerically, with the result $\zeta = 0.208 298 04
\epsilon +0.006858 \epsilon ^{2}$. This compares well with numerical
simulations, see figure \ref{fig:numstat}.

\begin{figure}\centerline{\small
\begin{tabular}{|c|c|c|c|c|}
\hline
$\zeta _{\rm}$ & one loop & two loop & estimate & 
simulation and exact\\
\hline
\hline
$d=3$  & 0.208 &  0.215  & $0.215\pm 0.01$  & 
$0.22\pm 0.01$ \cite{Middleton1995}  \\
\hline
$d=2$ &0.417 &0.444 &$0.42\pm 0.02$ &  $0.41\pm 0.01$ \cite{Middleton1995} \\
\hline
$d=1$ & 0.625 & 0.687 &  $0.67\pm 0.02$ & $2/3$ \\
\hline
\end{tabular}}\medskip 
\caption{Results for $\zeta $ in the random bond case.}\label{fig:numstat}
\end{figure}

\section{Large $N$}\label{largeN} In the last section, we have
discussed renormalization in a loop expansion, i.e.\ expansion in
$\E$. In order to independently check consistency it is good to have a
non-perturbative approach. This is achieved by the large-$N$ limit,
which can be solved analytically and to which we turn now. We start
from
\begin{eqnarray}\label{HlargeN}
{\cal H}[\vec u,\vec j ] &=& \frac{1}{2T} \sum _{a=1}^{n}\int_{x} 
 \vec u_{a} (x)\left(-\nabla^{2}{+}m^{2} \right) \vec u_{a} (x) - \sum
_{a=1}^{n}\int_{x} \vec{j}_{a} (x)\vec{u}_{a} (x)  \nn \\
&&   -\frac{1}{2 T^{2}}  \sum
_{a,b=1}^{n} \int_x B \left((\vec u_{a} (x)-\vec u_{b} (x))^{2} \right)\ .
\end{eqnarray}
where in contrast to (\ref{H}), we use an $N$-component field $\vec{u}
$. For $N=1$, we identify $B (u^{2} )=R (u)$. We also have added a
mass $m$ to regularize the theory in the infra-red and a source
$\vec{j} $ to calculate the effective action $\Gamma (\vec u) $ via a
Legendre transform. For large $N$ the saddle point equation reads
\cite{LeDoussalWiese2001}
\begin{equation}\label{saddlepointequation}
\tilde B' (u_{ab}^{2}) = B' \left(u_{ab}^{2}+2 T I_{1} + 4
I_{2} [\tilde B' (u_{ab}^{2})-\tilde B' (0)] \right)
\end{equation}
This equation gives the derivative of the effective (renormalized)
disorder $\tilde B$ as a function of the (constant) background field
$u_{ab}^{2}= (u_{a}-u_{b})^{2}$ in terms of: the derivative of the
microscopic (bare) disorder $B$, the temperature $T$ and the integrals
$I_{n}:= \int_{k}\frac{1}{\left(k^{2}+m^{2} \right)^{n}}$.

The saddle-point equation can again be turned into a closed functional
renormalization group equation for $\tilde B$ by taking the derivative
w.r.t.\ $m$:
\begin{equation}\hspace{-0.9 cm}
\partial _{l}\tilde B (x)\equiv -\frac{m \partial }{\partial m}\tilde
B (x) =\left(\epsilon -4\zeta \right)\! \tilde B (x) + 2 \zeta x
\tilde B' (x)+\frac{1}{2}\tilde B' (x)^{2}-\tilde B' (x) \tilde B'
(0)+ \frac{\epsilon\, T \tilde B' (x)}{\epsilon +\tilde B'' (0)}\,\,\,
\end{equation}
This is a complicated nonlinear partial differential equation. It is
therefore surprising, that one can find an analytic solution. (The
trick is to write down the flow-equation for the inverse function of
$\tilde B' (x)$, which is linear.) Let us only give the results of
this analytic solution: First of all, for long-range correlated
disorder of the form $\tilde B' (x)\sim x^{-\gamma }$, the exponent
$\zeta $ can be calculated analytically as $\zeta =\frac{\epsilon }{2
(1+\gamma )}\ . $ It agrees with the replica-treatment in
\cite{MezardParisi1991} and the 1-loop treatment in
\cite{BalentsDSFisher1993}. For short-range correlated disorder,
$\zeta =0$.  Second, it demonstrates that before the Larkin-length,
$\tilde B (x)$ is analytic and thus dimensional reduction
holds. Beyond the Larkin length, $\tilde B'' (0)=\infty $, a cusp
appears and dimensional reduction is incorrect. This shows again that
the cusp is not an artifact of the perturbative expansion, but an
important property even of the exact solution of the problem (here in
the limit of large $N$).

\section{Relation to Replica Symmetry Breaking (RSB)}\label{s:RSB} There
is another treatment of the limit of large $N$ given by M\'ezard and
Parisi \cite{MezardParisi1991}. They start from (\ref{HlargeN}) but
{\em without}\/ a source-term $j$. In the limit of large $N$, a
Gaussian variational ansatz of the form
\begin{eqnarray}\label{HlargeNMP}
{\cal H}_{\mathrm g}[\vec u] &=& \frac{1}{2T} \sum _{a=1}^{n}\int_{x} 
 \vec u_{a} (x)\left(-\nabla^{2}{+}m^{2} \right) \vec u_{a} (x) 
   -\frac{1}{2 T^{2}}  \sum
_{a,b=1}^{n} \sigma_{ab} \, \vec u_{a} (x)\vec u_{b} (x)
\end{eqnarray}
becomes exact. The art is to make an appropriate ansatz for
$\sigma_{ab}$. The simplest possibility, $\sigma _{ab}=\sigma $ for
all $a\neq b$ reproduces the dimensional reduction result, which
breaks down at the Larkin length. Beyond that scale, a replica
symmetry broken (RSB) ansatz for $\sigma _{ab}$ is suggestive. To this
aim, one can break $\sigma _{ab} $ into four blocks of equal size,
choose one (variationally optimized) value for the both outer diagonal
blocks, and then iterate the procedure on the diagonal blocks,
resulting in
\begin{equation}\label{RSB}
\sigma_{ab} =
\left(\,\parbox{.25\textwidth}{\fig{.25\textwidth}{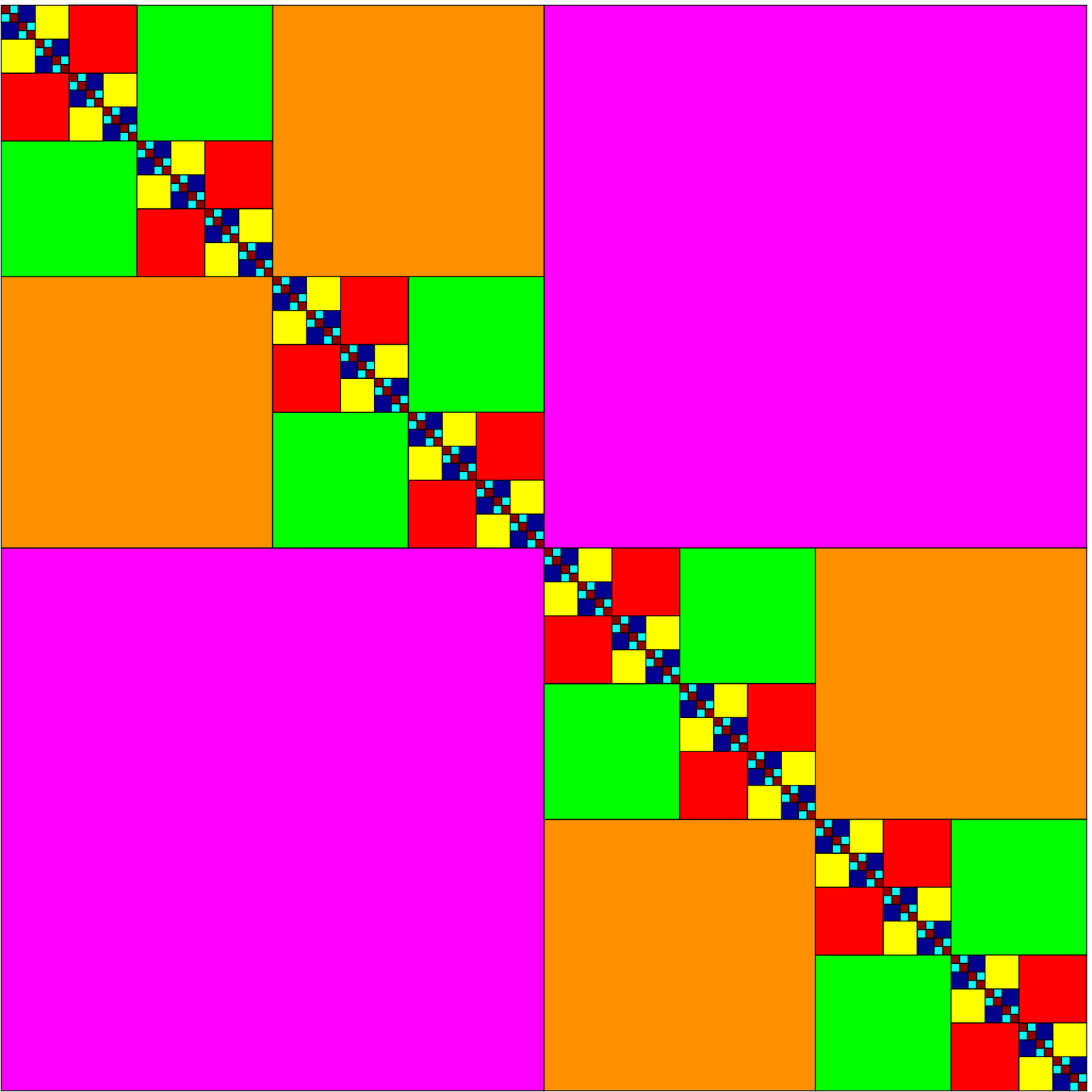}}\,\right)\ .
\end{equation}\begin{figure}[b]
\centerline{\fig{8cm}{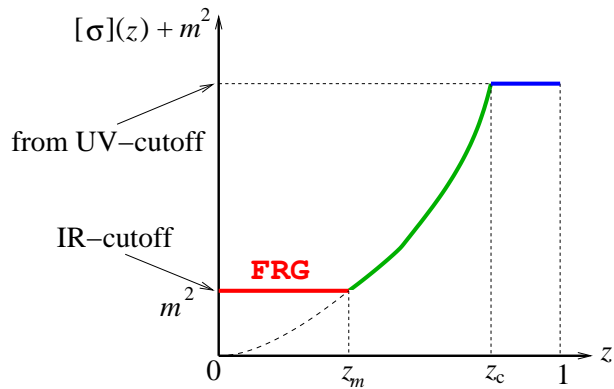}}
\caption{The function $\left[\sigma \right] (u)+m^{2}$ as given in
\protect\cite{MezardParisi1991}.} \vspace{-0.1cm}\label{fig:MP-function}
\end{figure} 
One finds that the more often one iterates, the better the results. In
fact, one has to repeat this procedure infinite many times. This seems
like a hopeless endeavor, but Parisi has shown that the infinitely
often replica symmetry broken matrix can be parameterized by a
function $[\sigma] (z)$ with $z\in \left[0,1 \right]$. In the
SK-model, $z$ has the interpretation of an overlap between
replicas. While there is no such simple interpretation for the model
(\ref{HlargeNMP}), we retain that $z=0$ describes distant states,
whereas $z=1$ describes nearby states. The solution of the large-$N$
saddle-point equations leads to the curve depicted in figure 6.
Knowing it, the 2-point function is given by
\begin{equation}\label{RSBformula}
\left< u_{k}u_{-k} \right>=\frac{1}{k^{2}}\left(1+\int_{0}^{1} \frac{\rmd
z}{z^{2}}  \frac{\left[\sigma  \right] (z)+m^{2}}{k^{2}+\left[\sigma  \right] (z)+m^{2}} \right)\ .
\end{equation}
The important question is: What is the relation between the two
approaches, which both pretend to calculate the same 2-point function?
Comparing the analytical solutions, we find that the 2-point function
given by FRG is the same as that of RSB, if in the latter expression
we only take into account the contribution from the most distant
states, i.e.\ those for $z$ between 0 and $z_{m}$ (see figure
\ref{fig:MP-function}). To understand why this is so, we have to
remember that the two calculations were done under quite different
assumptions: In contrast to the RSB-calculation, the FRG-approach
calculated the partition function in presence of an external field
$j$, which was then used to give via a Legendre transformation the
effective action. Even if the field $j$ is finally turned to 0, the
system might remember its preparation, as is the case for a magnet:
Preparing the system in presence of a magnetic field will result in a
magnetization which aligns with this field. The magnetization will
remain, even if finally the field is turned off. The same phenomena
happens here: By explicitly breaking the replica-symmetry through an
applied field, all replicas will settle in distant states, and the
close states from the Parisi-function $\left[\sigma \right] (z)+m^{2}$
(which describes {\em spontaneous} RSB) will not contribute.  However,
we found that the full RSB-result can be reconstructed by remarking
that the part of the curve between $z_{m}$ and $z_{c}$ is independent
of the infrared cutoff $m$, and then integrating over $m$
\cite{LeDoussalWiese2001} ($m_{c}$ is the mass corresponding to
$z_{c}$):
\begin{equation}\label{RSB=intFRG}
\left< u_{k}u_{-k} \right>\Big|^{\mathrm{RSB}}_{k=0} =\frac{\tilde
R'_{m}(0)}{m^{4}} +\int_{m}^{m_{c}} \frac{\rmd \tilde
R'_{\mu}(0)}{\mu^{4}} + \frac{1}{m_{c}^{2}}-\frac{1}{m^{2}}\ .
\end{equation}
We also note that a similar effective action  has been proposed in
\cite{BalentsBouchaudMezard1996}. While it agrees qualitatively, it
does not reproduce the correct FRG 2-point function, as
it should.

\section{Finite $N$}\label{s:finiteN} Up to now, we have studied the
functional RG in two cases: For one component $N=1$ and in the limit
of a large number of components, $N\to \infty$. The general case of
finite $N$ is more difficult to handle, since derivatives of the
renormalized disorder now depend on the direction, in which this
derivative are taken. Define amplitude $u:=|\vec u|$ and direction
$\hat u:= \vec u/|\vec u|$ of the field. Then deriving the latter
variable leads to terms proportional to $1/u$, which are diverging in
the limit of $u\to 0$. This poses additional problems in the
calculation, and it is a priori not clear that the theory at $N\neq1$
exists, supposed this is the case for $N=1$. At 1-loop order
everything is well-defined \cite{BalentsDSFisher1993}. We have found a
consistent RG-equation at 2-loop order \cite{LeDoussalWiesePREPd}:
\begin{figure}[b] 
\centerline{{\unitlength1mm
\begin{picture} (90,55)
\put(0,0){\fig{85mm}{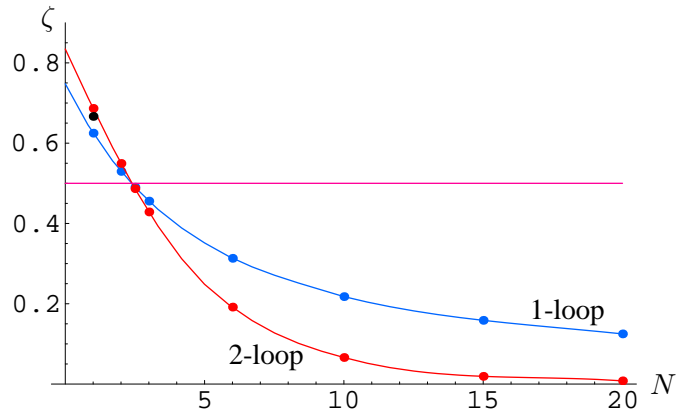}}
\put(5,52){$\zeta $}
\put(86,3){$N$} 
\put(70,13){1-loop}
\put(30,7){2-loop}
\end{picture}}
} 
\caption{Results for the roughness $\zeta$ at 1- and 2-loop order, as
a function of the number of components $N$.}  \label{f:Ncomp}
\end{figure}
\begin{eqnarray}\label{2loopFPENcomp}
\partial_{\ell } R(u) &=& (\epsilon - 4 \zeta) R(u) + \zeta u R'(u)
+\frac{1}{2} R''(u)^2 - R''(0) R''(u) +\frac{N-1}{2} \frac{R'(u)}{u}
\left(\frac{R'(u)}{u} - 2 R''(0)\right)
\nn \\
&&+\frac{1}{2} \left( R''(u) - R''(0) \right) \,{R''' (u)}^2
+\frac{N{-}1}{2} \frac{{\left( R'(u) {-} uR''(u) \right) }^2\, ( 2
R'(u) {+} u(R''(u) {-}3 R''(0) ) )}{u^5}
\nonumber \\
&&  -R'''(0^{+})^{2} \left[\frac{N+3}{8}R''(u)+\frac{N-1}{4}\frac{R'(r)}{u} \right]
\ .
\end{eqnarray}
The first line is the 1-loop equation, given in
\cite{BalentsDSFisher1993}. The second and third line represent the
2-loop equation, with   the new anomalous terms proportional to $R'''
(0^{+})^{2}$ (third line).

The fixed point equation (\ref{2loopFPENcomp}) can be integrated
numerically, order by order in $\epsilon$. The result, specialized to
directed polymers, i.e.\ $\epsilon =3$ is plotted on figure
\ref{f:Ncomp}.  We see that the 2-loop corrections are rather big at
large $N$, so some doubt on the applicability of the latter down to
$\epsilon=3$ is advised. However both 1- and 2-loop results reproduce
well the two known points on the curve: $\zeta =2/3$ for $N=1$ and
$\zeta =0$ for $N=\infty$. The latter result has been given in section
\ref{largeN}. Via the equivalence \cite{KPZ} of the directed polymer
problem in $N$ dimensions treated here and the KPZ-equation of
non-linear surface growth in $N$ dimensions, which relate the
roughness exponent $\zeta$ of the directed polymer to the dynamic
exponent $z_{\mathrm{KPZ}}$ in the KPZ-equation via $\zeta
=\frac{1}{z_{{\mathrm{KPZ}}}}$, we know that $\zeta (N=1)=2/3$.

The line $\zeta =1/2$ given on figure \ref{f:Ncomp} plays a special
role: In the presence of thermal fluctuations, we expect the
roughness-exponent of the directed polymer to be bounded by $\zeta \ge
1/2$. In the KPZ-equation, this corresponds to a dynamic exponent
$z_{\mathrm{KPZ}}=2$, which via the exact scaling relation
$z_{\mathrm{KPZ}}+\zeta_{\mathrm{KPZ}}=2$ is an upper bound in the
strong-coupling phase. The above data thus strongly suggest that there
exists an upper critical dimension in the KPZ-problem, with
$d_{\mathrm{uc}}\approx 2.4$. Even though the latter value might be
an underestimation, it is hard to imagine what can go wrong {\em
qualitatively} with this scenario. The strongest objections will
probably arize from numerical simulations, such as
\cite{MarinariPagnaniParisi2000}. However the latter use a discrete
RSOS model, and the exponents are measured for interfaces, which in
large dimensions have the thickness of the discretization size,
suggestions that the data are far from the asymptotic regime. We thus
strongly incourage better numerical simulations on a continuous model,
in order to settle this issue.

\section{Depinning transition}\label{s:dynamics}
\begin{figure}[b]
\centerline{\fig{0.4\textwidth}{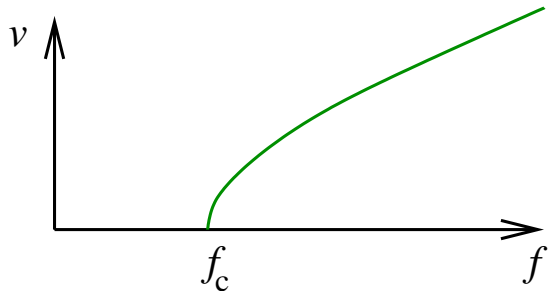}}
\caption{Velocity of a pinned interface as a function of the applied 
force. Zero force: equilibrium. $f=f_{c}$: depinning.}
\label{f:vel-force}
\end{figure}
Another important class of phenomena for elastic manifolds in disorder
is the so-called ``depinning transition'': Applying a constant force
to the elastic manifold, e.g.\ a constant magnetic field to the
ferromagnet mentioned in the introduction, the latter will only move,
if a certain critical threshhold force $f_{c}$ is surpassed, see
figure \ref{f:vel-force}. (This is fortunate, since otherwise the
magnetic domain walls in the hard-disc drive onto which this article
is stored would move with the effect of deleting all information,
depriving the reader from his reading.)  At $f=f_{c}$, the so-called
depinning transition, the manifold has a distinctly different
roughness exponent $\zeta$ (see Eq.~(\ref{roughness})) from the
equilibrium ($f=0$). For $f>f_{c}$, the manifold moves, and close to
the transition, new observables and corresponding exponents appear:
\begin{itemize}
\itemsep0mm
\item  the dynamic exponent $z$ relating correlation functions in
spatial and temporal direction
$$
t\sim x^{\,z}
$$
\item a correlation length $\xi$ set by the distance to $f_{c}$
$$
\xi \sim |f-f_{c}|^{-\nu }
$$
\item  furthermore, the new exponents are not all independent, but
satisfy the following exponent relations
\cite{NattermanStepanowTangLeschhorn1992}
\begin{equation}\label{exp-relatons}
\beta =\nu (z- \zeta ) \qquad \qquad \nu =\frac{1}{2-\zeta }
\end{equation}
\end{itemize}
The equation describing the movement of the interface is
\begin{equation}\label{eq-motion}
\partial_{t} u (x,t) = (\nabla^{2}+m^{2}) u (x,t) + F (u (x,t),x) \ ,
\qquad F (u,x)=-\partial_{u} V (u,x)
\end{equation}
This model has been treated at 1-loop order by Natterman et
al.~\cite{NattermanStepanowTangLeschhorn1992} and by Narayan and
Fisher \cite{NarayanDSFisher1993a}. The 1-loop flow-equations are
identical to those of the statics. This is surprising, since
physically, the phenomena at equilibrium and at depinning are quite
different. There is even the claim by \cite{NarayanDSFisher1993a},
that the roughness exponent in the random field universality class is
exactly $\zeta =\epsilon /3$, as it is in the equilibrium random field
class. After a long debate among numerical physicists, the issue is
today resolved: The roughness is significantly larger, and reads e.g.\
for the driven polymer $\zeta =1.25$, instead of $\zeta=1$ as
predicted in \cite{NarayanDSFisher1993a}. Clearly, a 2-loop analysis
\cite{LeDoussalWieseChauve2002} is necessary, to resolve these
issues. Such a treatment starts from the dynamic action
\begin{equation}\label{dyn-action}
{\cal S} = \int_{x,t} \tilde u (x,t) (\partial_{t}-\nabla^{2}+m^{2}) u
(x,t) +\int_{x,t,t'} \tilde u (x,t)\Delta (u (x,t)-u (x,t'))\tilde u
(x,t')\ ,
\end{equation}
where the ``response field'' $\tilde u (x,t)$ enforces the equation of
motion (\ref{eq-motion}) and 
\begin{equation}\label{Delta}
\overline{F (u,x) F (u',x')} = \Delta (u-u')\delta^{d} (x-x') \equiv
-R'' (u-u') \delta^{d}(x-x')
\end{equation}
is the force-force correlator, leading to the second term in
(\ref{dyn-action}). As in the statics, one encounters terms
proportional to $\Delta' (0^{+})\equiv -R''' (0^{+})$. Here the
sign-problem can uniquely be resolved by supposing that the membrane
only jumps ahead,
\begin{equation}\label{jump-ahead}
t>t'\qquad \Rightarrow \qquad u (x,t)\ge u (x,t')\ .
\end{equation}
Practically this means that when evaluating diagrams containing
$\Delta (u (x,t)-u (x,t'))$, one splits them into two pieces, one with
$t<t'$ and one with $t>t'$. Both pieces are well defined, even in the
limit of $t\to t'$. As the only tread-off of this method, diagrams can
become complicated and difficult to evaluate; however they are always
well-defined.

\begin{figure}[b]
\scalebox{1.0}{
\begin{tabular}{|c|c|c|c|c|r|}
\hline 
 & $d$ & $\epsilon$ & $\epsilon^2$ & estimate & simulation~~~\\
\hline 
\hline 
        & $3$ & 0.33 & 0.38 & 0.38$\pm$0.02 & 0.34$\pm$0.01  \\
\hline 
$\zeta$ & $2$ & 0.67 & 0.86 & 0.82$\pm$0.1 & 0.75$\pm$0.02  \\
\hline 
        & $1$ & 1.00 & 1.43 & 1.2$\pm$0.2 & 1.25$\pm$0.01   \\
\hline 
\hline 
        & $3$ & 0.89 & 0.85 & 0.84$\pm$0.01 & 0.84$\pm$0.02  \\
\hline 
$\beta$ & $2$ & 0.78 & 0.62 & 0.53$\pm$0.15  & 0.64$\pm$0.02 
\\
\hline 
        & $1$ & 0.67 & 0.31 & 0.2$\pm$0.2 &  0.25 \dots 0.4  \\
\hline 
\hline 
        & $3$ & 0.58 & 0.61 &   0.62$\pm$0.01  & \\
\hline 
$\nu$ & $2$ & 0.67 & 0.77 &  0.85$\pm$0.1   & 0.77$\pm$0.04  \\ 
\hline 
        & $1$ & 0.75 & 0.98 &   1.25$\pm$0.3  & 1$\pm$0.05  \\ 
\hline
\end{tabular}}\hfill 
%
{%
\begin{tabular}{|c|c|c|c|c|c|}
\hline 
 & $\epsilon $ & $\epsilon ^{2}$ & estimate & simulation \\
\hline 
\hline 
$\zeta $& 0.33     &  0.47    & 0.47$\pm$0.1 & 0.39$\pm$0.002 \\
 \hline
 $\beta $  & 0.78 & 0.59 & 0.6$\pm $0.2 &0.68$\pm$0.06   \\
\hline
$z$  &0.78 &0.66 &0.7$\pm $0.1 & 0.74$\pm$0.03 \\
\hline $\nu $ & 1.33 & 1.58 & 2$\pm$0.4 &1.52$\pm $0.02 \\
\hline
\end{tabular}} 
\caption{The critical exponents at the depinning transition, for short
range elasticity (left) and for long range elasticity (right).}
\label{dyn-data}
\end{figure}
Physically, this means that we approach the depinning transition from
above. This is reflected in (\ref{jump-ahead}) by the fact that $u
(x,t)$ may remain constant; and indeed correlation-functions at the
depinning transition are completely independent of
time\cite{LeDoussalWieseChauve2002}.  On the other hand a theory for
the approach of the depinning transition from below ($f<f_{c}$) has
been elusive so far.

At the depinning transition, the 2-loop functional RG reads
\cite{ChauveLeDoussalWiese2000a,LeDoussalWieseChauve2002}
\begin{eqnarray}\label{two-loop-FRG-dyn}
\partial_{\ell} R (u) \!&=&\! (\epsilon -4 \zeta)R (u)+\zeta u R' (u)
+\frac{1}{2}R'' (u)^{2} {-}R'' (u) R'' (0) \nonumber \\
&&+\frac{1}{2}\,\left[R'' (u)-R'' (0) \right] R''' (u)^{2}
~{\mbox{\bf +}}~ \frac{1}{2}\, R''' (0^{+})^{2} R''
(u)
\end{eqnarray}
First of all, note that it is a priori not clear that the functional
RG equation, which is a flow equation for $\Delta (u)=-R'' (u)$ can be
integrated to a functional RG-equation for $R (u)$. We have chosen
this representation here, in order to make the difference to the
statics evident: The only change is in the last sign on the second
line of (\ref{two-loop-FRG-dyn}). This has important consequences for
the physics: First of all, the roughness exponent $\zeta$ for the
random-field universality class changes from $\zeta
=\frac{\epsilon}{3}$ to
\begin{equation}\label{zetaRFdyn}
\zeta =\frac{\epsilon}{3} (1 +0.14331 \epsilon +\dotsb )
\end{equation}
Second, the random-bond universality class is unstable and always
renormalizes to the random-field universality class, as is physically
expected: Since the membrane only jumps ahead, it always experiences a
new disorder configuration, and there is no way to know of wether this
disorder can be derived from a potential or not. This non-potentiality
is most strikingly observed in the random periodic universality class,
which is the relevannt one for charge density waves. The fixed point
for a periodic disorder of period one reads (remember $\Delta (u)=-R''
(u)$)
\begin{equation}\label{rand-per-fp}
\Delta^{*} (u) =\frac{\epsilon}{36}+\frac{\epsilon^{2}}{108}
-\left(\frac{\epsilon}{6}+\frac{\epsilon^{2}}{9} \right) u (1-u)
\end{equation}
Integrating over a period, we find (suppressing in $F (u,x)$ the
dependence on the coordinate $x$ for simplicity of notation)
\begin{equation}\label{period}
\int_{0}^{1}\rmd u \, \Delta^{*} (u) \equiv \int_{0}^{1}\rmd u\ 
\overline{F (u) F (u')}= -\frac{\epsilon^{2}}{108}\ .
\end{equation}
In an equilibrium situation, this correlator would vanish, since
potentiality requires $\int_0^{1}\rmd u\, F (u)\equiv 0$. Here, there
are non-trivial contributions at 2-loop order (order $\epsilon^{2}$),
violating this condition and rendering the problem non-potential. This
same mechanism is also responsible for the violation of the conjecture
$\zeta =\frac{\epsilon}{3}$, which could be proven on the assumption
that the problem remains potential under renormalization. Let us
stress that the breaking of potentiality under renormalization is a
quite novel observation here.

The other critical exponents mentioned above can also be
calculated. The dynamical exponent $z$ (for RF-disorder) reads
\cite{ChauveLeDoussalWiese2000a,LeDoussalWieseChauve2002}
\begin{equation}\label{zdyn}
z=2-\frac{2}{9}\epsilon -0.04321\epsilon^{2} + \dotsb 
\end{equation}
All other exponents are related via the relation (\ref{exp-relatons}).
That the method works well even quantitatively can be infered from
figure \ref{dyn-data}.  \section{Universal width
distribution}\label{s:distribution}
\begin{figure}[b] \centerline{\fig{0.5\textwidth}{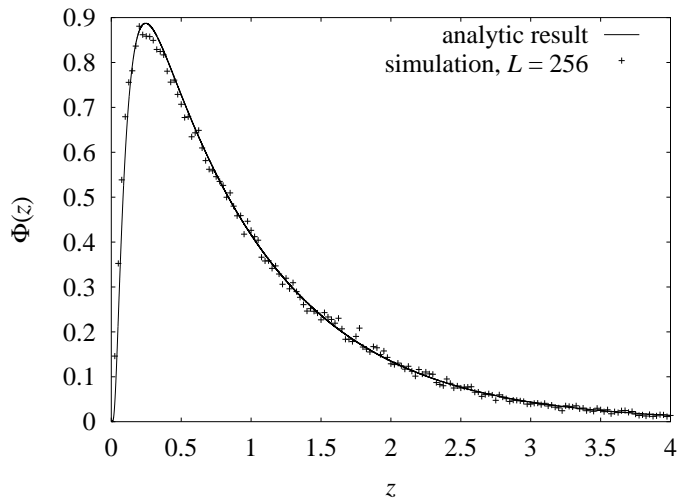}}
\caption{Scaling function $\Phi(z)$ for the ($1+1$)--dimensional
harmonic model, compared to the Gaussian approximation for
$\zeta=1.25$. Data from
\protect\cite{RossoKrauthLeDoussalVannimenusWiese2003}.}
\label{f:Dhm}
\end{figure}
Exponents are not the only interesting quantities: In experiments and
simulations, often whole distributions can be measured, as e.g.\ the
universal width distribution of an interface that we have computed at
depinning
\cite{RossoKrauthLeDoussalVannimenusWiese2003,LeDoussalWiese2003a}. Be
$\left< u \right>$ the average position of an interface for a {\em
given} disorder configuration, then the spatially averaged width
\begin{eqnarray}\label{w2}
w^{2}:= \frac{1}{L^{d}}\int_{x}\left(u (x)-\left< u \right> \right)^{2}
\end{eqnarray}
is a random variable, and we can try to calculate and measure its
distribution $P (w^{2})$. The rescaled function $\Phi (z)$, defined by
\begin{equation}\label{Phi}
P (w^{2})={1}/{\overline{w^{2}}}\,\Phi
\left({w^{2}}/{\overline{w^{2}}} \right)
\end{equation}
will be universal, i.e.~independent of microscopic details and the
size of the system.

Supposing all correlations to be Gaussian, $\Phi (z)$ can be
calculated analytically. It depends on two parameters, the roughness
exponent $\zeta$ and the dimension $d$. Numerical simulations
displayed on figure \ref{f:Dhm} show spectacular agreement between
analytical and numerical results. As expected, the Gaussian
approximation is not exact, but to see deviations in a simulation,
about $10^{5}$ samples have to be used. Analytically, corrections can
be calculated: They are of order $R''' (0^{+})^{4}$ and
small. Physically, the distribution is narrower than a Gaussian.


\section{Anisotropic depinning, directed percolation, branching and
all that}\label{s:anisotopic}
\begin{figure}[b] \centerline{\fig{2cm}{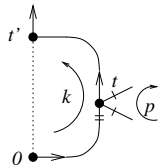}} \caption{{The
diagram generating the irreversible nonlinear KPZ term with one
disorder vertex and one $c_4$ vertex (the bars denote spatial
derivatives).}}  \label{fig1.a}
\end{figure} 
We have discussed in section \ref{s:dynamics} isotopic depinning,
which as the name suggest is a situation, where the system is
invariant under a tilt. This isotropy can be broken through an
additional anharmonic elasticity 
\begin{equation}\label{Eanharm}
E_{\mathrm{elastic}}= \int_{x} \frac{1}{2}\left[\nabla
u (x)\right]^{2}+c_{4} \left[\nabla u (x)\right]^{4}\ ,
\end{equation}
leading to a drastically different universality class, the so-called
anisotropic depinning universality class, as found recently in
numerical simulations \cite{RossoKrauth2001b}. It has been observed in
simulations \cite{AmaralBarabasiStanley1994,TangKardarDhar1995}, that
the drift-velocity of an interface is increased, which can be
intepreted as a tilt-dependent term, leading to the equation of motion
in the form
\begin{equation}\label{lf28}
 \partial_t u (x,t)= \nabla^2 u (x,t) + \lambda \left[ \nabla u
(x,t)\right]^2+ F(x,u (x,t) ) + f\ .
\end{equation}
However it was for a long time unclear, how this new term
(proportional to $\lambda$), which usually is refered to as a
KPZ-term, is generated, especially in the limit of {\em vanishing}
drift-velocity. In  \cite{LeDoussalWiese2002a} we have shown that this
is possible in a non-analytic theory, due to the diagram given in
figure \ref{fig1.a}.

For anisotropic depinning, numerical simulations based on cellular
automaton models which are believed to be in the same universality
class
\cite{TangLeschhorn1992,BuldyrevBarabasiCasertaHavlinStanleyVicsek1992},
indicate a roughness exponent $\zeta \approx 0.63$ in $d=1$ and $\zeta
\approx 0.48$ in $d=2$.  On a phenomenological level it has been
argued
\cite{TangLeschhorn1992,BuldyrevBarabasiCasertaHavlinStanleyVicsek1992,GlotzerGyureSciortinoConiglioStanley1994}
that configurations at depinning can be mapped onto directed
percolation in $d=1+1$ dimensions, which yields indeed a roughness
exponent $\zeta_{\mathrm{DP}}= \nu_\perp/\nu_{\|} = 0.630 \pm 0.001$,
and it would be intriguing to understand this from a systematic field
theory.

This theory was developed in \cite{LeDoussalWiese2002a}, and we review
the main resulsts here. A strong simplification is obtained by going
to the Cole-Hopf transformed fields
\begin{equation}\label{lf29}
Z ( {x,t}) := \rme^{ \lambda u(x,t)} \qquad \Leftrightarrow \qquad
u(x,t) = \frac{\ln ( Z(x,t))}{ \lambda } \ .
\end{equation}
The equation of motion becomes after multiplying with $ \lambda
Z(x,t)$ (dropping the term proportional to $f$)
\begin{equation}\label{lf30}
   \partial_t Z(x,t) = \nabla^2 Z(x,t) +{{\lambda} }
F\left(x,\frac{\ln (Z(x,t))}{{\lambda} } \right) Z(x,t)
\end{equation}
and the dynamical action (after averaging over disorder)
\begin{equation}\label{cole}
{\cal S} = \int_{xt}\tilde {Z}(x,t)\left( \partial_{t}-\nabla^{2}
\right) Z(x,t) -\frac{ \lambda^{2} }{2 } \int_{xtt'} \tilde {Z}(x,t)
{Z}(x,t) \, \Delta\! \left( \frac{\ln Z(x,t)-\ln Z(x,t')}{ \lambda
}\right)\tilde {Z}(x,t')Z(x,t')
\end{equation}
This leads  to the FRG flow equation at 1-loop order
\begin{eqnarray}
\partial _{\ell} \Delta (u) &=& (\epsilon -2\zeta )
\Delta (u) + \zeta u 
 \Delta' (u)  -\Delta'' (u)\left(
\Delta (u)-\Delta (0) \right) - \Delta' (u)^2\nn \\
&&+2 \lambda \Delta (u) \Delta' (0^{+}) +2 \lambda ^{2}\left(\Delta
(u)^{2} +\Delta (u)\Delta (0)\right) \label{beta-2}
\end{eqnarray}
The first line is indeed equivalent to (\ref{1loopRG}) using $\Delta
(u)=-R'' (u)$. The second line is new and contains the terms induced
by the KPZ term, i.e.\ the term proportional to $\lambda$ in
(\ref{lf28}).

Equation (\ref{beta-2}) possesses the following remarkable property:
{\em A three parameter subspace of exponential functions forms an
exactly invariant subspace.}  Even more strikingly, this is true {\it
to all orders} in perturbation theory \cite{LeDoussalWiese2002a}! The
subspace in question is ($0\le u\le 1/\lambda $)
\begin{equation}\label{lf80}
\Delta(u) = \frac{\epsilon }{ \lambda^2} \left(a + b\, \rme^{- \lambda
u} + c\, \rme^{\lambda u}\right)
\end{equation}
\begin{figure*}[!t] \centerline{\fig{.5\textwidth}{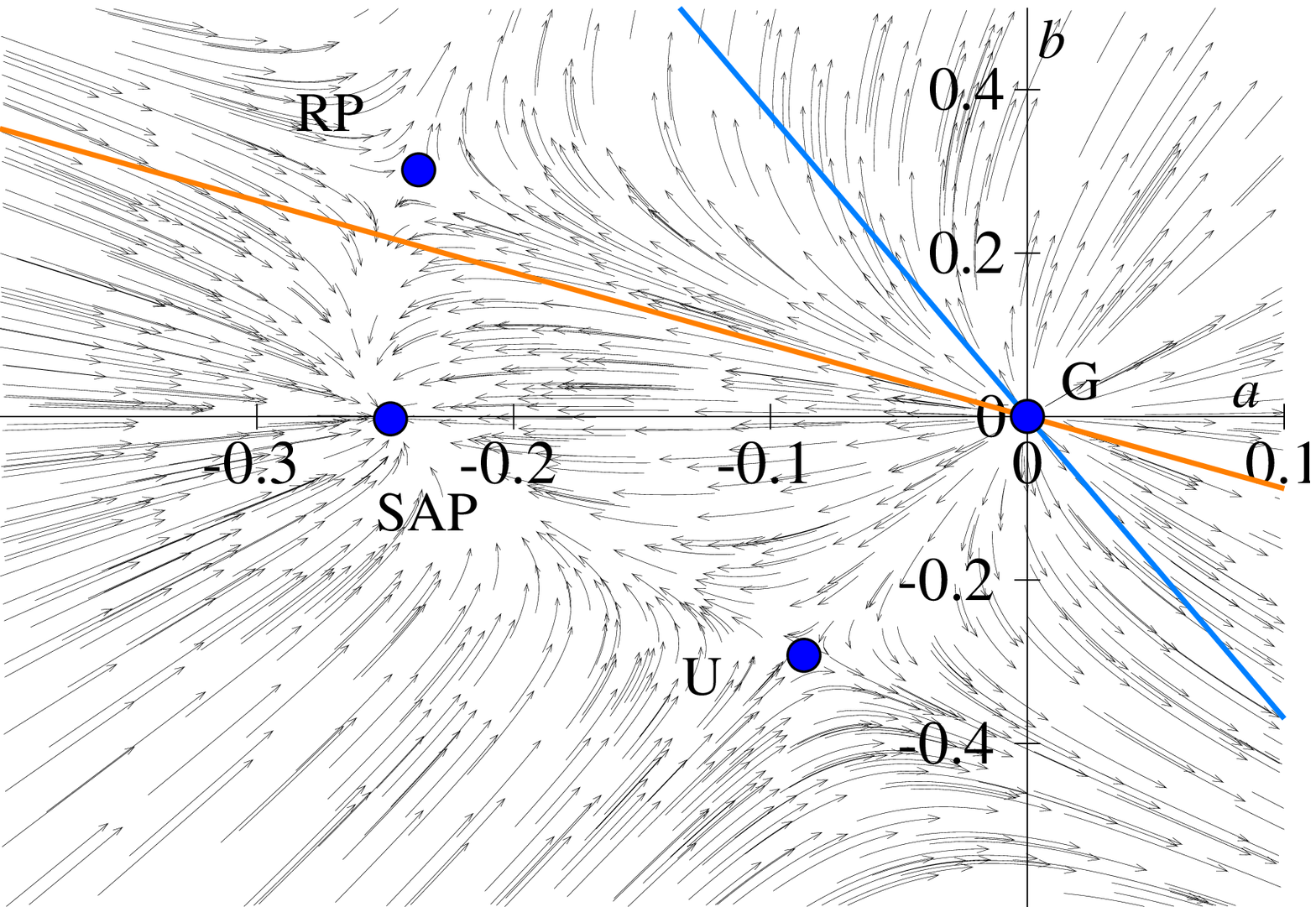}}
\caption{Fixed point structure for $\lambda=2$, which is a typical
value. The ratio $c/b$ is not renormalized, see
(\ref{lf43})-(\ref{lf44}), such that $c/b$ is a parameter, fixed by
the boundary conditions, especially $\lambda $. The fixed points are
Gaussian {\tt G}, Random Periodic {\tt RP} (the generalization of the
RP fixed point for $\lambda =0$), Self-Avoiding Polymers {\tt SAP},
and Unphysical {\tt U}.}  \label{lambda-flow}
\end{figure*}%
The FRG-flow (\ref{beta-2}) closes in this subspace,
leading to the simpler 3-dimensional flow:
\begin{eqnarray}
 \partial_{\ell} a &=& a + 4 a^2 + 4 a c + 4 b c \label{lf42}\\
 \partial_{\ell} b &=& b (1 + 6 a  + b + 5 c ) \label{lf43}\\
 \partial_{\ell} c &=& c (1 + 6 a  + b + 5 c )\label{lf44}
\end{eqnarray} This flow has a couple of fixed points, given on figure
\ref{lambda-flow}. They describe different physical situations. The
only globally attractive fixed point is {\tt SAP}, describing
self-avoiding polymers. This fixed point is not attainable from the
physically relevant inital conditions, which lie (as fixed point {\tt
RP}) between the two separatrices given on figure \ref{lambda-flow}.
All other fixed points are cross-over fixed points.

\begin{figure}[b] \centerline{\fig{0.5\textwidth}{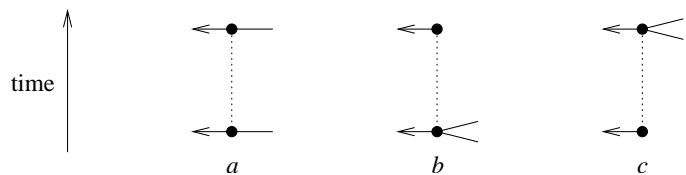}}
\caption{The three vertices proportional to $a$, $b$ and $c$ in
equation (\ref{c2}).}  \label{f:branch}
\end{figure}In the Cole-Hopf representation, it is easy to see why the
exponential manifold is preserved to all orders. Let us insert
(\ref{lf80}) in (\ref{cole}).  The complicated functional disorder
takes a very simple polynomial form \cite{LeDoussalWiese2002a}.
\begin{equation}\label{c2}
{\cal S}=\int_{xt}\tilde {Z}(x,t)\left( 
\partial_{t}-\nabla^{2}  \right) Z(x,t)- \int_{x}\int_{t<t'}
\tilde {Z}(x,t)\tilde {Z}(x,t') \left(a
Z(x,t)Z(x,t')+bZ(x,t)^{2}+cZ(x,t')^{2} \right)\ .  
\end{equation}
The vertices are plotted on figure \ref{f:branch}. It is intriguing to
interprete them as particle interaction ($a$) and as different
branching processes ($b$ and $c$): $Z$ destroys a particle and $\tilde
Z$ creates one. Vertex $b$ can e.g.\ be interpreted as two particles
coming together, annihilating one, except that the annihilated
particle is created again in the future. However, if the annihilation
process is strong enough, the reappearance of particles may not play a
role, such that the interpretation as particle annihilation or
equivalently directed percolation is indeed justified.

One caveat is in order, namely that the fixed points described above,
are all crossover fixed point, and nothing can a priori be said about
the strong coupling regime. However this is the regime seen in
numerical simulations, for which the conjecture about the equivalence
to directed percolation has been proposed. Thus albeit intriguing, the
above theory is only the starting point for a more complete
understanding of anisotropic depinning. Probably, one needs another
level of FRG, so as standard FRG is able to treat directed polymers,
or equivalently the KPZ-equation in the absence of disorder.

\section{Perspectives}\label{perspectives} More interesting problems
have been treated by the above methods, and much more remains to be
done.  We have applied our techniques to the statics at 3-loop order
\cite{LeDoussalWiesePREPb} and to the random field problem
\cite{LeDoussalWiesePREPf}.  An expansion in $1/N$, (by now we have
obtained the effective action \cite{LeDoussalWiesePREPc}), should
allow to finally describe such notorious problems as the
strong-coupling phase of the Kardar-Parisi-Zhang equation. Many
questions are still open. Some have already been raised in these
notes, another is wether FRG can also be applied to spin-glasses. We
have to leave this problem for future research and as a challenge for
the reader to plunge deeper into the mysteries of functional
renormalization.

\subsection*{Acknowledgements}\label{ack} It is a pleasure to thank
the organizers of TH 2002 for the opportunity to give this
lecture. The results presented here have been obtained in a series of
inspiring ongoing collaborations with Pierre Le Doussal, and I am grateful to
him and my other collaborators Pascal Chauve, Werner Krauth and  Alberto
Rosso for all their enthousiasm and dedicated work. Numerous
discussions with Leon Balents, Edouard Br\'ezin and Andreas Ludwig are
gratefully acknowledged.


\end{document}